\documentclass[fleqn,9pt,twocolumn]{wlscirep}
\usepackage[utf8]{inputenc}
\usepackage[T1]{fontenc}
\usepackage{amsmath,amssymb}
\usepackage{mathtools}
\usepackage{arydshln}
\usepackage[font=small,labelfont=bf,
   justification=justified,
   format=plain, figurename=Fig.]{caption}

\title{Transition of Social Organisations Driven by Gift Relationship}
\author[1]{Kenji Itao}
\author[2, 3, *]{Kunihiko Kaneko} 

\affil[1]{Department of Basic Science, Graduate School of Arts and Sciences, University of Tokyo, Komaba 3-8-1, Meguro-ku, Tokyo 153-8902, Japan.}
\affil[2]{Research Center for Complex Systems Biology, University of Tokyo.}
\affil[3]{The Niels Bohr Institute, University of Copenhagen, Blegdamsvej 17, Copenhagen, 2100-DK, Denmark.}

\affil[*]{kaneko@complex.c.u-tokyo.ac.jp}

\begin{abstract}
Anthropologists have observed gift relationships that establish social relations as well as the transference of goods in many human societies. The totality of such social relations constitutes the network. Social scientists have analysed different types of social organisations with their characteristic networks. However, the factors and mechanisms that cause the transition between these types have hardly been explained. Here, we focus on the gift as the driving force for such changes. We build the model by idealising gift interactions and simulating the consequent social change due to long-term massive interactions. We demonstrate the emergence of disparities and various social organisations depending on the frequency of the gift, consistent with the empirical data. The constructive simulation study, as presented here, explains how people's interactions shape various social structures in response to environmental conditions. Combined with empirical studies, this could contribute to the formulation of a general theory in the social sciences.\end{abstract}
\begin{document}

\flushbottom
\maketitle
\thispagestyle{empty}

Gift relations accompanying the establishment of social relations and transference of goods have been observed worldwide.
In societies where the gift is important, three obligations of giving, receiving, and reciprocating arise \cite{mauss1923essai}. Particularly, reciprocation unites people, whereas those who fail to reciprocate lose their reputations and become subordinate to the donor. Other researchers, in contrast, emphasize that some gifts are donated to seek acknowledgement without expecting reciprocation \cite{moriyama2021gift}. In any case, the gift strengthens the social relations between the donor and recipient, including cooperation, dominance, and subordination.

The totality of social relationships constitutes the network \cite{leach1982social, wasserman1994social}. Different social organisation structures exist for such networks. 
Human beings generally form kinship systems based on genealogical and marital relationships due to developed kin recognition \cite{levi1969elementary, chapais2009primeval, planer2021towards, rand2013human}.
As the population density and the frequency of conflicts with external enemies increase, the structure shifts from bands united by kinship to tribes united by collective ideas such as brotherhood, to chiefdoms composed of role-divided groups, and then to states with legitimate monopolies of power \cite{service1962primitive}. Furthermore, the increase in economic and social disparities accompanies these transitions. Other researchers have emphasized the increase in productivity, surplus, or frequency of war to explain the emergence of complex social organisations, disparities, and the division of labour \cite{marx1911contribution,bataille1949part, carneiro1970theory, peter2009evolution}. However, their origins have been unclear.

In this paper, we focus on the gift relationship that causes a change in microscopic interpersonal relations as the driving force that shapes the macroscopic structure of social organisations. Therefore, we model the gift relations. People transfer their assets, produce them, and reciprocate for the gift. When reciprocation succeeds, an equal cooperative relation is established. However, when it fails, the recipient would repay for reciprocation and be subordinate to the donor.
By simulating the model, we demonstrate the emergence of various social organisations. We show that social structure shifts from bands to tribes and chiefdoms, depending on the frequency of the gift. Furthermore, we demonstrate that as the gift transactions are more frequent, economic disparity followed by social disparity arises.
Thus, we bridge the gift theory of microscopic interpersonal relations and the theory of macroscopic social structures and provide the so-called ``mechanism-based explanation'' to reveal the micro-macro relation and historical causality \cite{ylikoski2012micro}.

To test the validity of the theoretical results, we compare them with the statistical analysis of a global ethnographic database of premodern societies called the standard cross-cultural sample (SCCS) \cite{murdock1969standard, kirby2016d}. The SCCS contains 186 societies, considered culturally and linguistically independent of each other (even if some correlation exists due to shared ancestry in the strict sense \cite{minocher2019explaining}). The data allow us to quantitatively analyse cultural adaptations to environments \cite{marsh1967comparative, bernard2017research}.
Subsequently, we empirically unveil the successive rise of economic and social disparities as the frequency of the gift increases.
By collaboration between theoretical simulation and statistical analysis, we produce logically coherent and empirically valid scenarios on the evolution of each social organisation.

The remainder of this paper is organised as follows: in the next section, we introduce a basic model of the gift relationship and demonstrate the emergence of economic and social disparities. Following this, we extend the model to include the generation alternation to consider kin relations and demonstrate the transition of social organisations from bands to tribes and chiefdoms.
Next, by analysing the SCCS, the theoretical results are verified. Finally, we discuss the significance of this method, which combines theoretical models with empirical data analysis, to explore social phenomena.

\section*{Basic model}
\subsection*{Model}
\begin{figure}[tb]
\centering
\includegraphics[width= 0.7\linewidth]{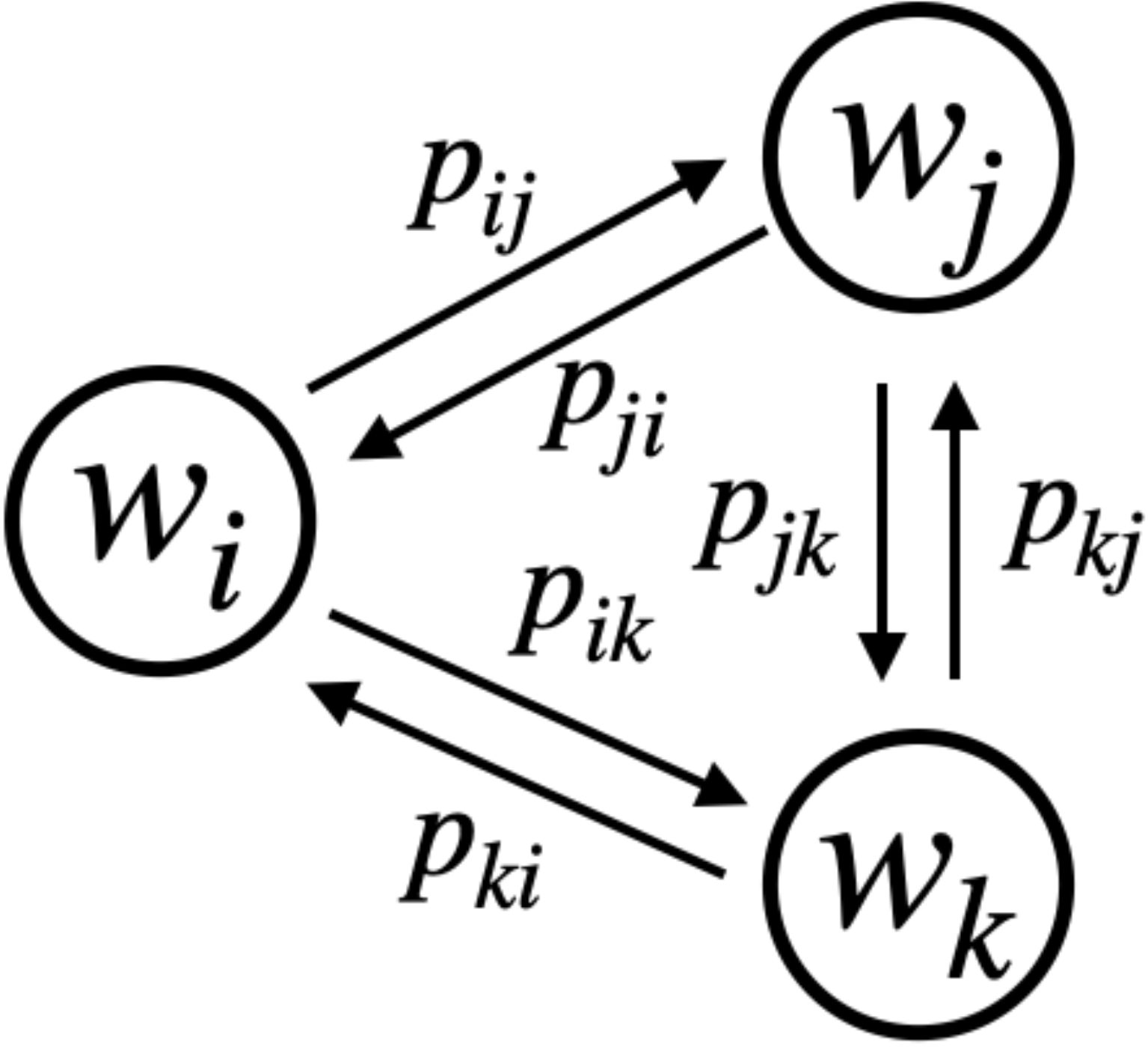}
\caption{Schematic of the model.
Each node represents the individual. Node $i$ has its own wealth $w_i$. The weight of directed edge $p_{ij}$ represents the probability of individual $i$ giving a gift to $j$.
}
\label{fig:gift_model}
\end{figure}

First, as a preparation, we introduce a basic model for the development of economic and social relations by gift within a single generation.
In the model, we represent society as a network of people. The nodes represent individuals, and the directed edges represent their social relations as shown in Fig. \ref{fig:gift_model}. Each node $i$ has its own wealth $w_i$. In each time step, each node $i$ gives its wealth as a gift to node $j$ with probability $p_{ij}$. Then, each node produces the wealth. The production of node $i$ is proportional to $1 + \log(1 + w_i)$, considering the law of diminishing returns \cite{malthus1798essay, ricardo1891principles, gibson2011land}. After production, each node reciprocates for the gift. Here, one must return $r$ times the amount received in the initial gift. When one can reciprocate appropriately, the deal ends. However, failure to reciprocate appropriately will result in a ``debt,'' which will be repaid based on the subsequent production. Until the repayment is completed, those repaying cannot make new gifts. Here, we assume that each node gives its entire wealth as an initial gift. However, the results are qualitatively unchanged even when each node gives some percentage, if not all, of its wealth as an initial gift.

Note that ``wealth'' here refers to any assets in the broad sense. It can include money, livestock, ornaments, and, in extreme cases, people. Such assets will increase productivity directly or indirectly since they may be factors of production or provide prestige for collecting such factors.
Wealth can be produced by any form of labour, including hunting, agriculture, herding, and city labour, as long as the wealth increases the productivity, although the increment may depend on the labour form. Strictly, the productivity functions depend on such labour forms, but here we neglect its dependence for simplicity by regarding wealth as a coarse-grained quantity.

To represent the change in social relations caused by the gift, we assume that the edge weight $p_{ij}$ increases by $\eta$ each time the wealth is transferred from node $i$ to $j$, whether as a gift, reciprocation, or repayment. The more frequently one gives to a person, the greater the motivation to make a gift to that person in the future or the greater the awareness of social relations oriented toward that person. Appropriate reciprocation results in $p_{ij} \simeq p_{ji} > \eta$, that is, $i$ and $j$ are in an equal cooperative relationship. However, repaying the reciprocation over several steps by $i$ for the gift from $j$ will lead to $p_{ij} >> p_{ji}$, that is, $j$ has an advantage over $i$. Note that the edge weights are normalized so that $\sum_{j} p_{ij} = 1$ after the change in their values due to the gift, to fix the sum of the gift probability to $1$.

In the simulation, the initial edge weights are set equal to $1 / N$, where $N$ is the number of nodes. Similarly, each node has the wealth $1.0$ in the initial state. The parameters are summarized in Table. \ref{table:params}

\begin{table}[tb]
\caption{Parameters used in the model. In the results described below, parameter values are fixed to those shown in the table, unless the value is described explicitly. The last two parameters below the dashed line only appear in the full model.
} 
\label{table:params}
\centering
\begin{tabular}{lll} 
Sign	&	Explanation	& Value	\\\hline
$N$	&	(Initial) number of individuals in society	&	$100$	\\
$\eta$	&	Increment of edge weight by transaction    &$0.03$	\\
$r$&	Interest rate for reciprocation		&$1.1$ (Variable)	\\\hdashline
$N_s$& Number of societies in the model &$100$\\
$l$	&	Average time of gift in a generation	& Variable	
\end{tabular}
\end{table}

\subsection*{Emergence of disparity}
\begin{figure*}[tb]
\centering
\includegraphics[width= \linewidth]{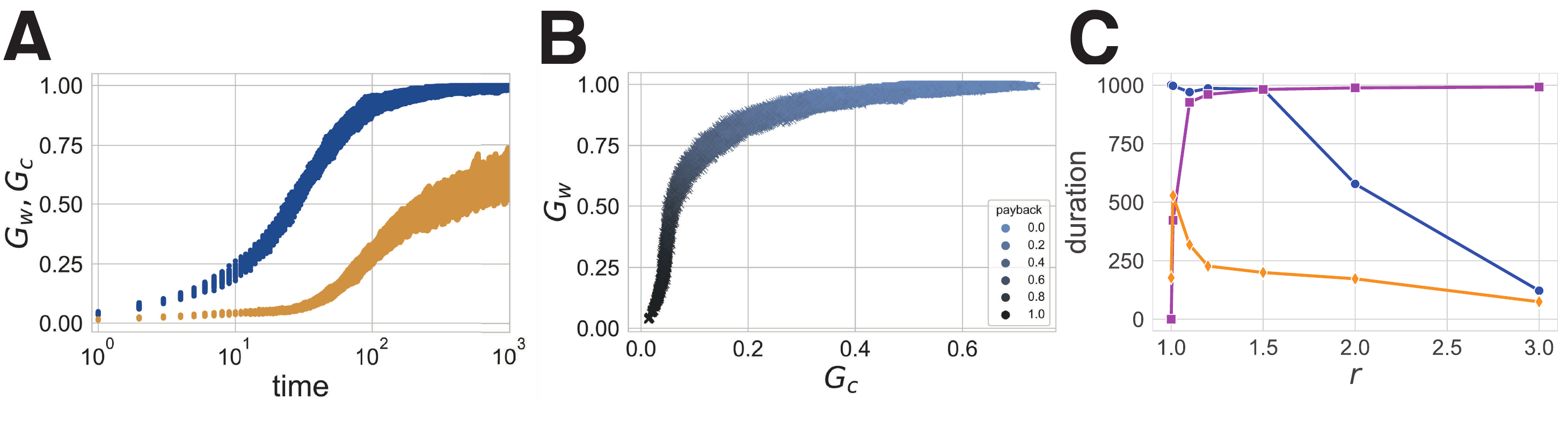}
\caption{Simulation result of the basic model. (A) Temporal change of Gini coefficients for wealth (blue) and connectivity (yellow). The graph presents the results of 100 simulations. (B) Scatter plots of Gini coefficients at each time step. Marker colour indicates the percentage of appropriate reciprocation made for the gift at that time. The lighter colour shows that fewer gifts are reciprocated. (C) Average duration of individuals being ``free'' (blue), ``repaying'' (purple), and ``rich'' (orange), depending on the interest rate for reciprocation $r$. ``Free'' indicates that people are not in the process of repayment, `repaying’’ indicates that they are in that process, and ``rich'' indicates that they are in the top 5\% of wealth in the society.
}
\label{fig:gift_basic_Gini}
\end{figure*}

Simulations are performed for 1,000 time steps, that is, 1,000 cycles of gift, production, and reciprocation procedures. With time, people's wealth and the network structures change. By analysing the distributions of wealth and connectivity, that is, the summation of edge weights directed to each node after each cycle, we observe the increase in the economic and social disparities. Fig. \ref{fig:gift_basic_Gini} (A) shows the temporal change of the Gini coefficients for wealth and connectivity. Gini coefficient for wealth $G_w$ is given by $G_w = \sum_{i = 1}^N \sum_{j=1}^N |w_i - w_j| / (2N\sum_{i = 1}^N w_i)$, which shows the extent of the inequality (the same applies to that of connectivity $G_c$).

Since gift transactions involve amplified reciprocation as long as $r > 1$, the economic disparity increases as more gifts are exchanged. When economic disparity is sufficiently large, the appropriate reciprocation becomes difficult. Then, unidirectional social relationships are established through the repayment of ``debts.''
As shown in Fig. \ref{fig:gift_basic_Gini} (A, B), the increase in economic disparity precedes social disparity. Furthermore, social disparity increases when more people cannot reciprocate appropriately.

In Fig. \ref{fig:gift_basic_Gini} (C), we plot the average duration of individuals' statuses by changing the interest rate $r$. We focus on three statuses, i.e., ``free'' (blue), ``repaying'' (purple), and ``rich'' (orange). ``Free'' indicates that people are not in the repayment process, ``repaying'' indicates that they are in that process, and ``rich'' indicates that they are in the top 5\% of the wealthy in society. We calculate the average steps for which people sustain these statuses. The graph suggests that people are likely to lose their positions of ``free'' or ``rich'' as $r$ is larger, that is, when they have to reciprocate to a more considerable degree. Of course, there is no disparity or stable rich people for $r = 1$.

Fig. S1 shows the temporal change of the network structures. As time passes, the networks are denser, and the edges are concentrated toward a few people. Consequently, we observe the emergence of social disparity and hierarchical organisation in this model. However, when the number of gift interactions is small, the network is sparse and exhibits no specific structure. In real societies, people recognize kinship as well as gift relations \cite{planer2021towards, rand2013human, service1962primitive}. The literature suggests the importance of both kinship and reciprocal transactions in establishing social relationships \cite{von2019dynamics, thomas2018kinship, apicella2012social}. Therefore, we need to implement the reproduction process to include kin relationships to explain the transition of social organisations from bands to tribes and to chiefdoms.

\section*{Full model}
\subsection*{Model}
In this section, we introduce the reproduction process to the basic model. First, we assigned a lifetime $l_i$ to each individual $i$ following the Poisson distribution with a mean of $l$, which represents the average times one makes a gift in a lifetime. When $l_i$ steps have passed since the $i$'s birth, the individual $i$ reproduces children who inherit their wealth and network. Children inherit an equal division of their parent's wealth and the edge weights directed toward the parent. When the individual $i$ has $N_i$ children, the wealth of the children would be $w_i / N_i$ and the edge weight from the individual $j$ to the children would be $p_{ji} / N_i$. Subsequently, to model the kin relationship, siblings are connected by the edge with $3\eta$ of the weight. Note that this value is not important. For example, we can obtain essentially identical results by setting it as $5\eta$ or $10\eta$.

Here, since the number of children in families is positively correlated with their wealth in pre-industrial societies \cite{gibson2011land}, we assume that the number of children for individual $i$ follows the Poisson distribution with a mean of $1 + \log(1 + w_i)$. However, the following results are almost qualitatively independent of the specific forms. For example, assuming that it follows the Poisson distribution with a mean of $2$, independently of wealth, we can obtain essentially identical results regarding the increase in disparities and the transition of network structures.

The number of individuals in society will change through reproduction. In real societies, large societies eventually divide \cite{service1962primitive}. In this model, if the population of each society doubles the initial value $N$, we assume that it splits into two. At this time, connections with those who have split into different societies are removed and connectivity is renormalized to keep the sum of edge weights directed from each node to $1$. In the model, $N_s$ societies exist. When a society splits into two, another society will be removed from the system at random to keep the number of societies to $N_s$. This process can be interpreted as an invasion, imitation, or coarse-grained description of a growing system. Therefore, societies that grow faster replace others, resulting in society-level evolution. This multilevel evolution of families and societies follows the hierarchical Moran process, which is generally applied to the evolution of the group-level structure in hierarchical systems \cite{itao2020evolution, itao2021evolution, itao2022emergence, takeuchi2017origin, traulsen2006evolution}.

\subsection*{Transition of social organisations}
\begin{figure*}[tb]
\centering
\includegraphics[width= \linewidth]{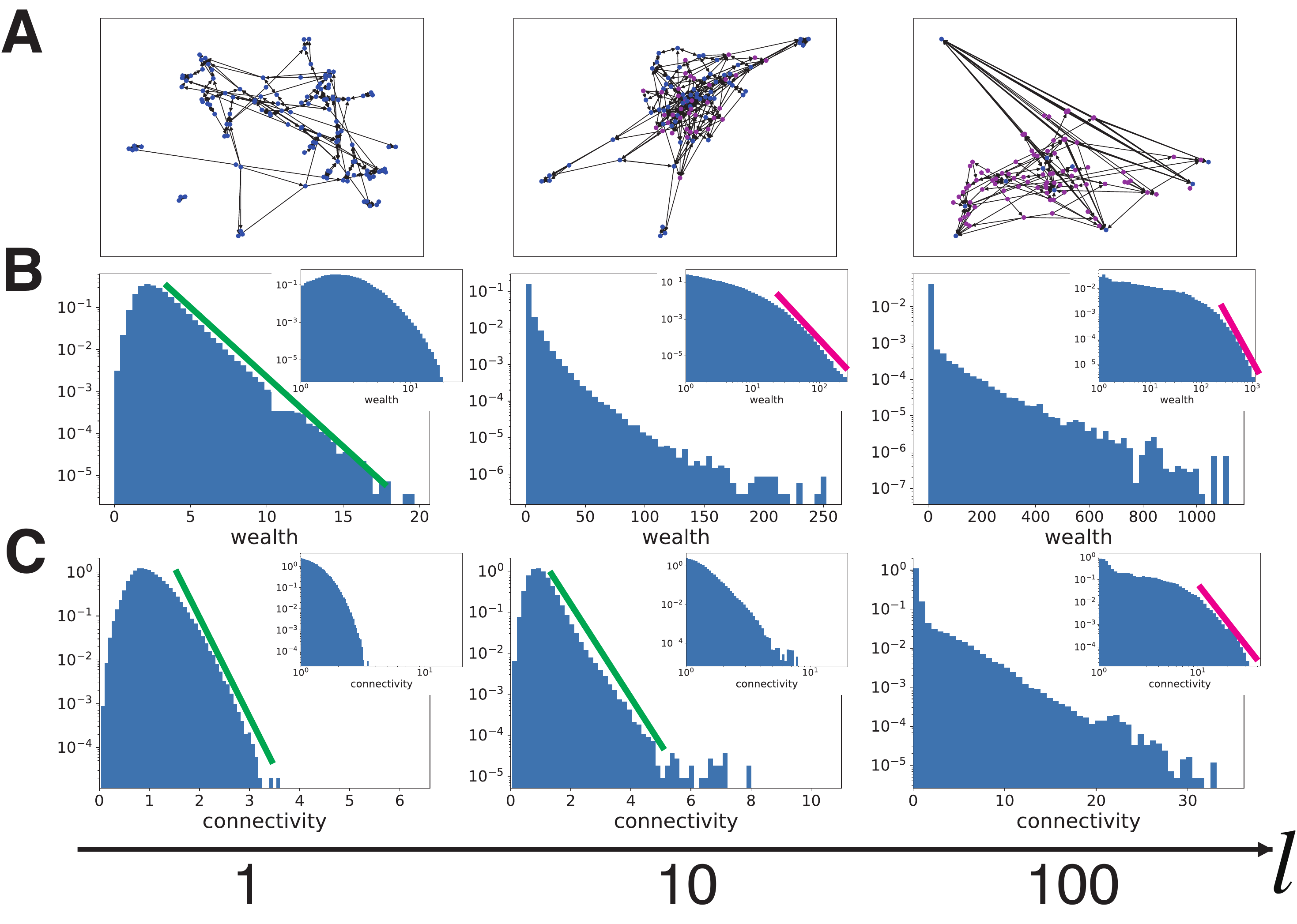}
\caption{Dependence of network structure (A), wealth distribution (B), and connectivity distribution (C) on the average lifetime $l$. (A) Network structure at the final state. Edges with weights larger than $\eta$ are shown. Red nodes represent ``repaying'' -- those who are in the repayment process. Blue nodes represent ``free,'' -- those who are not in that process. (B) The frequency distribution of wealth at the final state. (C) The frequency distribution of connectivity, i.e., the sum of the weights of edges directed to each node, at the final state. Insets are log-log plots of the frequency distributions. The green line is the exponential fitting of the distribution, whereas the pink lines are power-law fitting. The estimated exponents of power, that is, the slope of the lines for wealth distributions, are approximately 3 for $l = 10$ and 2 for $l = 100$. For the connectivity distribution, it is 4 for $l = 100$. The figures show the typical result for each of the three ``phases’’ described below. 
}
\label{fig:gift_full_result}
\end{figure*}

We performed the simulation for $100l$ steps by changing $l$ and $r$. Fig. \ref{fig:gift_full_result} shows the network structures and the distributions of wealth and connectivity of individuals in societies after $100l$ steps. Within this time, the system has converged to an approximately steady state. When $l$ is small, that is, the frequency of the gift is small, we observe small clusters of nodes that are densely connected to each other within the cluster and sparsely connected to the rest of the nodes.
Clusters are formed by shared ancestry that is, kinship within several generations. The gift relations bring sparse connection among clusters. At this time, the inequalities in wealth and connectivity are weak. As the average times of the gift in life $l$ increases, the clusters are larger and the connection among them is denser. Furthermore, there appears a strong inequality in wealth. Then, when $l$ is sufficiently large, the edges are concentrated on several ``free'' people who are not in the repayment process (blue nodes). The network is hierarchically organised with a chain of unidirectional edges. Both inequalities in wealth and connectivity are now strong. The distributions of wealth and connectivity show the power-law tail for the larger side as $w^{-\alpha}$ (or $c^{-\alpha}$). Such fat tails indicate strong disparities. Moreover, in Fig. \ref{fig:gift_full_result}(B) $\alpha$ is about 3 for $l = 10$ and 2 for $l = 100$. Such decrease in $\alpha$ values, i.e., the fatter tail for the larger side, indicates the further development of disparities for larger $l$.

\begin{figure*}[tb]
\centering
\includegraphics[width= 1.0\linewidth]{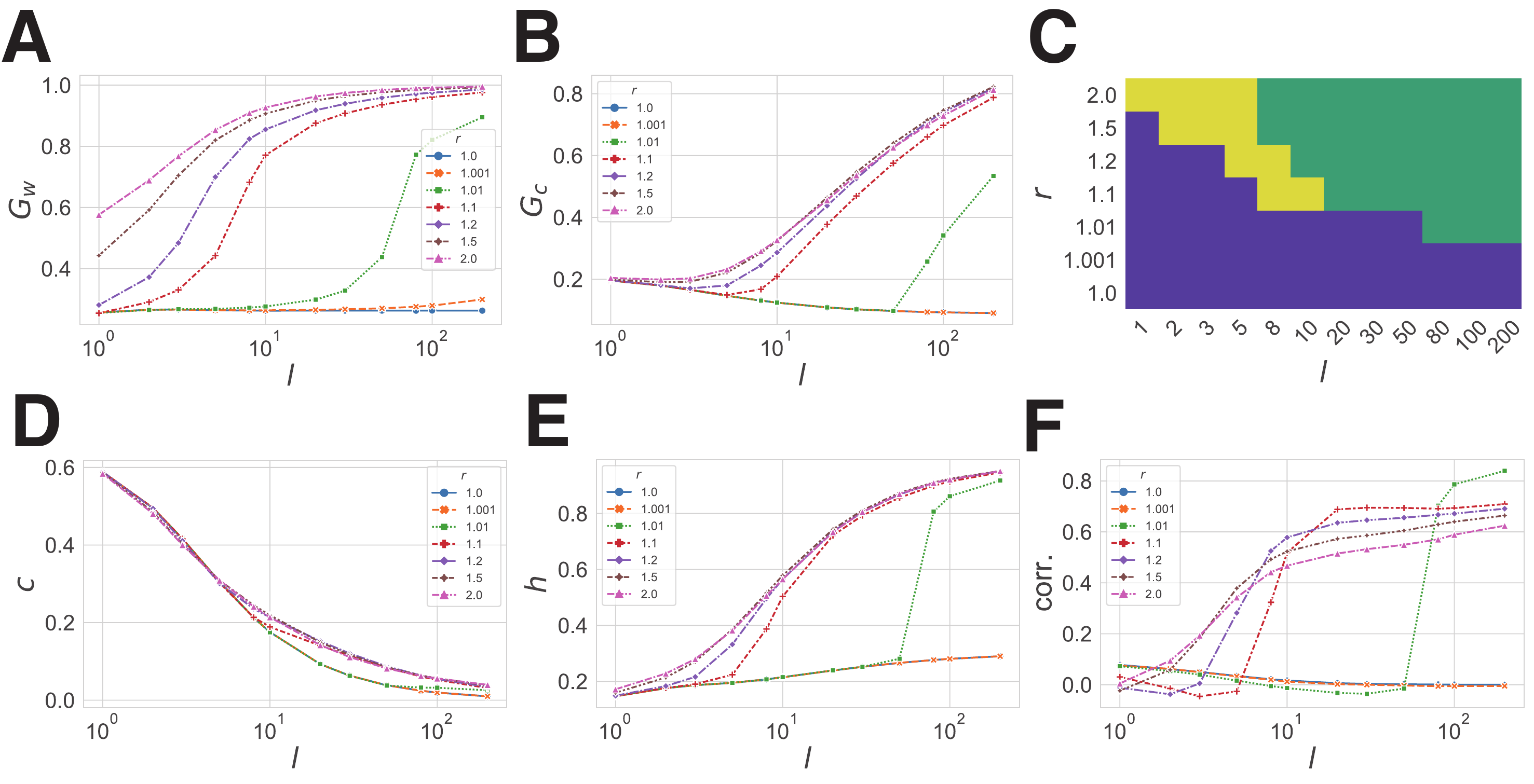}
\caption{Dependence of disparities and network characteristics on the average lifetime $l$ and interest rate $r$. (A) Gini coefficient for wealth $G_w$. (B) Gini coefficient for connectivity $G_c$. (C) Phase diagram of disparities. Phases of no disparities (blue), only economic disparity (yellow) and economic and social disparities (green) are shown. (D) Average clustering coefficient $c$, i.e., the cliquishness of a typical neighbourhood. (E) Flow hierarchy degree $h$, which is the persistent directionality in continuing flows. (F) Correlation coefficient of wealth and connectivity. Different line colours correspond to different interest rates $r$.
}
\label{fig:gift_full_index}
\end{figure*}

Fig. \ref{fig:gift_full_index}(A, B) reveals that the economic disparity arises before the social, as in the basic model\footnote{This successive emergence of power-law tail in the wealth and connectivity distribution recalls the embedding of power-law in the abundance of chemicals to that in reaction network connectivity \cite{furusawa2006evolutionary}.}. Furthermore, the disparities (and other quantities) for $r>1$ converge to those for $r = 1$ for small $l$ until they deviate sharply from those for $r = 1$.
This suggests that disparities emerge sharply, and the economic and social state moves to a different phase. Economic changes occur first, supposedly when the increase in production due to the acquisition of wealth is no longer sufficient to cover the interest for reciprocation. This is followed by social change, which is supposed to occur when most gifts are no longer reciprocated and the unidirectional social relations develop, as in the basic model.
Note that we have demonstrated the temporally successive emergence of economic and social disparities in the basic model. Here, however, we demonstrate their emergence as the adaptation to the different environmental parameter values $l$ and $r$.
Thus, we obtain the phase diagram on the disparities by examining whether each Gini coefficient doubles that for $r = 1$, as shown in Fig.\ref{fig:gift_full_index}(C). Parameter regions for no disparities, economic disparity only, and both disparities are shown in purple, yellow, and green, respectively, as distinct phases. The diagram shows that a longer average lifetime $l$ and a higher interest rate $r$ accelerate the evolution of disparities.

We then investigate the characteristics of emergent networks. Both social scientists and network theorists focus on the degree of clustering, that is, the cliquishness of a typical neighbourhood, and the degree of hierarchy, which is the asymmetric connectivity of different levels \cite{durkheim1982rules, wasserman1994social, watts1998collective, luo2011detecting}. The average clustering coefficient $c$ is measured by calculating the average percentage of the connection between each node's neighbours \cite{watts1998collective}.
Flow hierarchy degree $h$ is measured by calculating the percentage of edges that are not included in any cycle, which indicated the extent of persistent directionality in continuing flows \cite{luo2011detecting}. 

Fig. \ref{fig:gift_full_index} (D, E) shows the dependence of clustering coefficient $c$ and hierarchy degree $h$ on the average lifetime $l$ and the interest rate $r$. As $l$ is larger, implying gift transactions are more frequent, $c$ is smaller and $h$ is larger. The trend for $c$ is almost independent of the reciprocation rate $r$ but $h$ is dependent on it. Precisely, the increase in $h$ accompanies that of $G_c$. Hence, a hierarchical social organisation is a qualitative change in the system, which evolves together with social disparity. Kinship-based connection unites people who are genealogically close to each other and creates clusters. As social relations expand through gifts, kinship-based ties fade and people are linked to any other person in society, making the network less clustered. Then, when gift transactions are sufficiently frequent and many gifts are inappropriately reciprocated, a few rich people solely build novel relations. Those who cannot reciprocate are forced to repay and strengthen the relation directed to the donor of the gift. Then the hierarchy emerges in the network. People who receive repayment from many people are increasingly wealthy, which allows them to make gifts to many others (in fact, the correlation between wealth and connectivity is large only when societies are hierarchically organised, as shown in Fig. \ref{fig:gift_full_index}(F)). As a result, those with a large connectivity further enhance it. Hence, network development follows the so-called ``preferential attachment'', which is known to result in the power-law tail of connectivity \cite{barabasi1999emergence, dorogovtsev2000structure, krapivsky2000connectivity}.
Fig. \ref{fig:gift_full_result} (C) shows the power-law tail in the distribution of connectivity for larger $l$. 

In the simulation so far, we have assumed that people give their entire wealth as a gift and that children divide the inheritance equally. In real societies, however, the size of the gift and the distribution of the inheritance can differ \cite{malinowski1922argonauts, strathern1971rope, harrell1997human,todd1999diversite, todd2011origine, colleran2014farming, gibson2011land}. Therefore, we perform the simulation of an extended model in which the percentage of wealth to be donated and the inequality in inheritance evolve over generations. Here, we assume that these strategy parameters are transmitted from parents to children with slight variation through ``mutation,'' by referring to previous studies in cultural evolution \cite{cavalli1981cultural, creanza2017cultural}.
As shown in Fig. S2, people will spend most of their wealth on gifts when the average lifetime $l$ is sufficiently large. Furthermore, an equal inheritance evolves for small $l$ and an exclusive inheritance evolves for large $l$, which is consistent with the empirical study \cite{service1962primitive}. Note that the results regarding the emergence of disparities or transitions in network structures are qualitatively robust against this modification.

\section*{Empirical data analysis}
We empirically test our theoretical results on the successive rise of economic and social disparities and the transition of social organisations along with the increase in the frequency of the gift, by using the SCCS database \cite{murdock1969standard, kirby2016d}. First, we estimate the degree of the gift for each society using variables related to the frequency of events that accompany the gift-like transaction as follows: \textit{Compensation Demands, Taxation Paid to the Community, Degree of Marriage Celebration, Market Exchange within the Local Community, Tribute/ Taxation/ Expropriation}. Then, we estimate the economic and social disparities by \textit{Number of Rich People, Number of Poor, Number of Dispossessed} and \textit{Administrative Hierarchy, Social Stratification, Removal of Leaders Who Are Incompetent or Disliked}, respectively. We normalise the values of each variable to set the mean $0$ and the variance $1$. We also change the sign if necessary so that the higher values corresponded to a higher degree. For some societies, data for some variables are lacking. For the estimation, we average the available values. Then, we normalise each measure so that the minimum is $0$ and the maximum is $1$. 
These measures are rather qualitative compared to the measures used for the simulation, such as the Gini coefficients. However, considering the limitation in the available data, these measures are adopted to roughly estimate their relationship and examine the validity of the theoretical results. See the supplementary information for a further explanation of these variables.

\begin{table}[tb]
\caption{Correlation between SCCS variables and the estimated gift degree (excerpt). 
See Table S1 for further information.
} 
\label{table:gift_data_correlation_excerpt}
\centering
\begin{tabular}{lr} 
Variable	&	Corr.	\\\hline
Resource Base	&	0.63	\\
Societal Complexity	&	0.62	\\
Adults Herd Small Animals	&	0.62	\\
Metalworking	&	0.59	\\
Population Density	&	0.58	\\
Levels of Political Hierarchy	&	0.56	\\
Children hunt with adults	&	-0.54	\\
Political Role Differentiation	&	0.53	
\end{tabular}
\end{table}

By calculating the correlation between the SCCS variables and the estimated gift degree, we investigate cultural and environmental characteristics that can be related to the frequency of the gift. Table \ref{table:gift_data_correlation_excerpt} shows the variables with high correlations with the gift degree. The gift degree is suggested to be larger in societies with larger population density or richer resources. It is also suggested to be larger in herding societies and smaller in hunting societies. The difference depending on subsistence patterns will be due to the difference in the type of wealth. As we have mentioned, we use ``wealth'' in the broadest sense, and the parameter values of the frequency of gift $l$ (and the interest rate $r$) can depend on the type of wealth. $l$ and $r$ are large for societies exchanging livestock, since they are suitable for transportation and easy to increase.
Additionally, our analysis suggests that societies are hierarchically organised and people are specialised as the gift degree is larger.

\begin{figure}[tb]
\centering
\includegraphics[width= 1.0\linewidth]{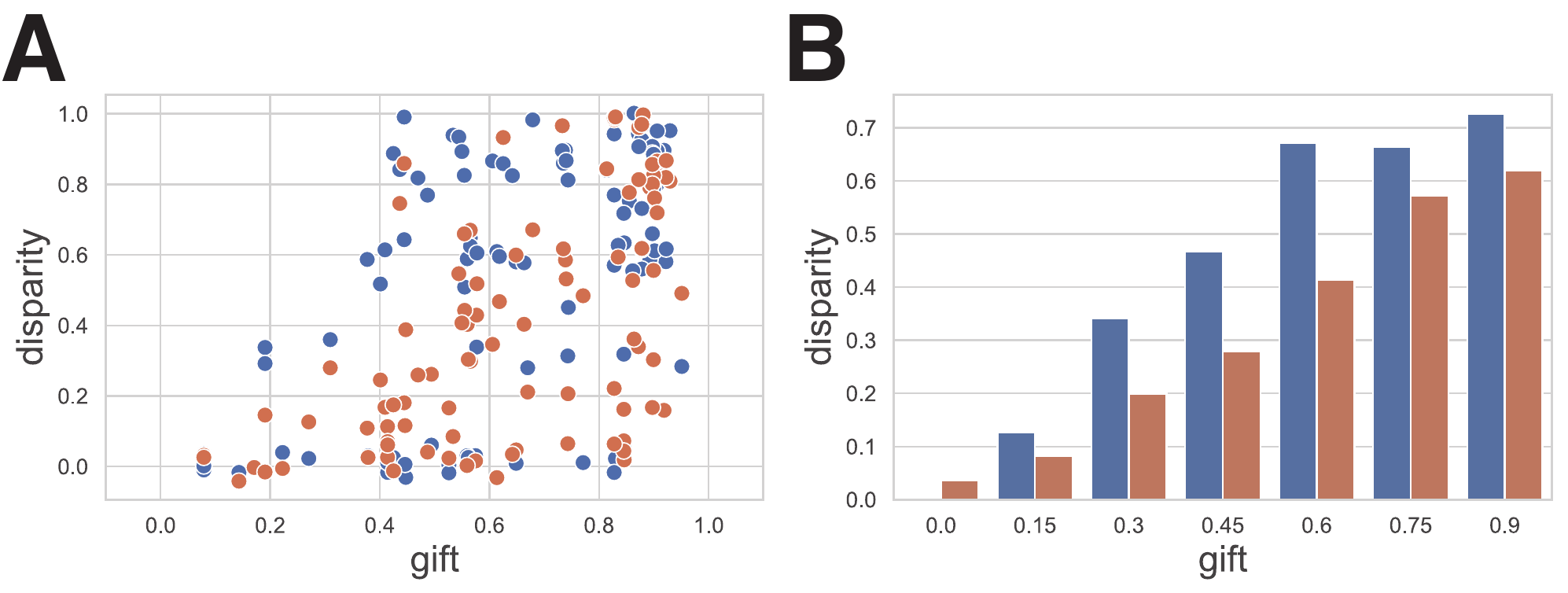}
\caption{Empirical relations between the gift degree and economic/ social disparities. (A) Scatter plots of the relation of economic (blue) and social (yellow) disparities to the gift degree. (B) Histogram corresponding to scatter plots.
}
\label{fig:gift_data_result}
\end{figure}

Fig. \ref{fig:gift_data_result} shows the empirical relations of economic and social disparities with the degree of gift. Consistent with the theoretical results in Figs. \ref{fig:gift_basic_Gini} and \ref{fig:gift_full_result}, both disparities increase with the gift degree. Furthermore, we also confirm that the increase in economic disparity precedes that of social disparity.

Unfortunately, the data on network features themselves are not available. However, the above correlation analysis is consistent with the emergence of hierarchical organisation under the large frequency of gift, which is predicted by the model. Furthermore, research on social organisations suggests that societies shift from band to tribe and to chiefdom as population density or the frequency of war increases \cite{service1962primitive}. Since a denser population and larger necessity of cooperation for war provide more opportunities for people to interact that include the gift, our theoretical results are consistent here.

Ethnographic reports suggest that the increase in tradable goods (often due to the contact with Westerners) enables more frequent exchange of gifts with largely amplified reciprocation. At this time, many people cannot maintain good status, while at the same time, great chiefs appear with economic and social dominance \cite{mauss1923essai, strathern1971rope}. This corresponds with the instability of statuses for a larger interest rate $r$ and a hierarchical organisation for larger frequency of the gift $l$ in our model.

\section*{Discussion}
By simulating the model of gift transactions, we demonstrated the emergence of disparities and the transition of social organisations. We found that societies shift among the three ``phases'' as the frequency and extent of the gift increase.
When gift transactions are infrequent, the kinship-based connection is dominant. People are equal, and society is composed of many small clusters of kin. As gift transactions occur frequently, economic disparity arises as a result of amplified reciprocation for the gift. Furthermore, people are densely connected so that larger clusters corresponding to tribes appear. Then, as transactions occur more frequently, many poor people fail to reciprocate, and social disparity arises due to the asymmetrical relation caused by the repayment of reciprocation. Societies are now hierarchically organised so that great chiefs appear with economic and social dominance. Then, we empirically verified these theoretical results through data analysis using the SCCS database. We confirmed that as gift transactions are more frequent, the economic and social disparities successively arise and societies are hierarchically organised.

Cultural anthropologists metaphorically interpret marriage as the ``gift (or exchange) of mates'' to emphasize that it brings the social relations to both partners' kin groups, including alliance and dominance, as comparable with the gift of goods \cite{levi1969elementary, leach1982social}. This, along with the genealogical relationship, has been considered the basic principle upon which human beings build kinship relationships \cite{leach1982social, planer2021towards}. Studies on kinship systems mainly focus on societies, such that these relationships are the main principles by which they are organised. (See \cite{itao2020evolution, itao2022emergence} for theoretical studies on the evolution of kinship structures therein.)

Increased population density and surplus production will accelerate the interaction of people, including the gift \cite{service1962primitive, bataille1949part}. A study suggests that the social network shrinks with the loss of surplus food \cite{von2019dynamics}. Our data analysis shows that the population density and richness of resources are positively correlated with the frequency of gift. Hence, it is suggested that the increase in population and productivity would accelerate the gift interaction, leading to the transition of social organisation from kinship systems. Note that our data analysis shows the correlation only. The above causality is suggested by the model but not empirically shown.

Our theoretical results suggest that as the frequency and scale of gift increases, economic disparity followed by social disparity arises. Economic disparity results from the amplified reciprocation for gift. When it is so large that most people cannot reciprocate, unidirectional relations are established, resulting in the emergence of social disparity and hierarchical organisation. At this time, the development of social networks follows the preferential attachment, and the power-law tail appears in the distribution of connectivity. The empirical results are consistent with such sequential emergence of disparities and hierarchical social organisation as the degree of gift increases. 

Furthermore, by comparing our results with Service's discussion, we note the correspondence between the phases in our model and his stages \cite{service1962primitive}. In his discussion, the band is characterised by small kin groups. In our model, the first phase is characterised by strongly clustered kin groups without economic or social disparities. The tribe is characterised by a large union of families with social equality. Our second phase is characterised by moderately clustered large groups with economic but not social disparities. Finally, the chiefdom is characterised by the hierarchical organisation of role-divided groups with both economic and social disparities. Our third phase is characterised by hierarchical network and its disparities.

Thus, our model would describe the following rough but logically coherent scenario for the development of human history: early in human history, the above ``gift of mates'' existed solely. Kinship structures were the dominant social organisations. Then the gift relations increase due to the generation of surplus through agriculture or pastoralism, the improvement of transportation, and the increase in population density. Consequently, social relations expand and social organisation shifts to tribes and then to chiefdoms. Additionally, people are specialised and societies are hierarchically organised. Here, we do not simply assume that surplus product can feed non-producers and allow the division of labour, but we see it as the driving force that promotes the gift, causing specialisation and social stratification. Previous studies have emphasised the increase in surplus, productivity, population density, and warfare \cite{service1962primitive, bataille1949part,marx1911contribution}. However, it was unclear how and why such factors cause social change. To solve such problems, it would be effective to perform simulations using a simple model, as we have presented here, to demonstrate such changes for providing the mechanistic explanation.

Here, it should be examined whether the gift is the main factor influencing people's social relations.
Anthropologists repeatedly observe the societies in which gift works as an important factor, especially in preindustrial societies including bands, tribes, and chiefdoms \cite{mauss1923essai, malinowski1922argonauts}. It is possible, however, that other factors may be more appropriate as driving factors of social changes, but this can only be evaluated by comparing our model with models built with a focus on other factors to determine which one of them explains the reality better.

The present study has some limitations. In the model, we ignore all other social relations except the gift. In reality, political or linguistic factors would also affect social relations. Some events may start or end the gift relationships, since gifts are made between people with shared values. Furthermore, our model ignores the intentional act of people. Indeed, such acts will guide social change, and we recognise its importance. Although we do not model them, by analogy with reference to the model, the historical facts would be better analysed. Here, we merely describe the statistical trend. This is where collaboration with ethnographers and historians is needed. 
Additionally, Service has proposed that phases of state with the legitimate monopoly of social power and industrial society with the complex and interdependent network of specialised groups appear after chiefdom \cite{service1962primitive}. Although the monopoly of wealth and network connectivity may be demonstrated in our model, we should consider other factors to discuss law enforcement or the balance of power between society and elites to reveal the evolution of states \cite{fukuyama2011origins, acemoglu2019narrow}. Sociologists have discussed that the change in attitude toward exchange and wealth precedes the emergence of industrial societies \cite{weber1930protestant, bataille1949part}. 
Hence, our model should be expanded to include these changes.

Our empirical data analysis also has some limitations. The estimation of the degrees of the gift and the economic or social disparities may seem arbitrary. To measure these variables directly, it is necessary to collaborate with field studies. Furthermore, we could only analyse the correlations between ethnographic variables and the gift degree.
Phylogenetic comparative analysis is also necessary to control statistical non-independence due to shared ancestry \cite{minocher2019explaining}. Additionally, because of the lack of chronological data, we could not analyse the causal relationships between the gift degree and disparities or social organisations.

Social structures are shaped through interactions among people over many generations. In this paper, we have theoretically demonstrated the formation of macroscopic social structures through microscopic interpersonal relations. We have built the model by idealising the behaviour of the people reported by anthropologists. Then, we examined the logical coherence of any macroscopic phenomenon if one assumes that many people behave that way. We find that the consequent macroscopic phenomena of the model are consistent with empirical observation.
By combining theoretical simulations of a simple constructive model and empirical data analysis, we have integrated the theory of interpersonal gift relations and that of social organisation, which have been discussed in a different context. Furthermore, we explain the origin of social organisations by revealing their microfoundation.
Theoretical studies, as shown here, produce explanatory scenarios by referring to empirical studies and propose relevant variables to be measured in the field. Empirical studies in the field describe notable phenomena and enable measurement of variables to test theories. This collaboration of theoretical and empirical studies will contribute to discussing the emergence of complex social structures and the unveiling of universal features in the social sciences.

\section*{Methods}
We adopted the following algorithm for the change in people's wealth and connectivity in the basic model. People randomly choose others to make gifts. Then, they produce wealth. Here, the productivity increases with the logarithm of their wealth. Finally, they reciprocate for the initial gift. For each transfer of wealth, the network connectivity, i.e., the weight of the edge, from the donor to the recipient is added by $\eta$.

A person $i$'s wealth $w_i$ and the edge weight from $i$ to $j$ $w_{ij}$ at the time $t$ are expressed as follows:
\begin{align}
\shortintertext{For each $i$, the recipient $j$ is chosen with the probability $p_{ij}^{t-1}$,}
w_{j}^{t\ast}  &= w_{j}^{t-1} + w_{i}^{t-1},  \label{eq:giving_wealth}\\
w_{i}^{t\ast} &= 0,  \label{eq:wealth_loss}\\
p_{ij}^{t\ast} &= p_{ij}^{t-1} + \eta, \label{eq:network_update1}\\
w_{i}^{t\ast\ast} &= w_{i}^{t\ast} + (1 + \log(1 + w_{i}^{t\ast})) / 100,  \label{eq:production}\\
\shortintertext{(note that $w_{j}^{t\ast}$ can be $> 0$ if $i$ receives a gift.)}
w_{i}^t  &= w_{i}^{t\ast\ast} + \min(w_{j}^{t\ast\ast}, rw_{i}^{t-1}),  \label{eq:reciprocation}\\
p_{ji}^{t\ast\ast} &= p_{ji}^{t\ast} + \eta, \label{eq:network_update2}\\
p_{ij}^t &= p_{ij}^{t\ast} / \sum_k p_{ik}^{t\ast}. \label{eq:connectivity_normalisation}
\end{align}

Each individual $i$ chooses the recipient $j$ according to the edge weights. $i$ gives the wealth to $j$ (Eqs. (\ref{eq:giving_wealth}, \ref{eq:wealth_loss})). Since wealth transfers from $i$ to $j$, the edge weight $p_{ij}$ is added by $\eta$ (\eqref{eq:network_update1}). Then, people produce their wealth (\eqref{eq:production}). Here, productivity increases with the logarithm of wealth. The denominator of productivity was set to $100$ to prevent wealth from exploding. However, its value can be any positive value to achieve qualitatively similar results. Next, people reciprocate for the initial gift (\eqref{eq:reciprocation}). If $j$'s wealth is larger than $r$ times the wealth received from $i$, i.e., $w_{i}^{t-1}$, $j$ reciprocate $rw_{i}^{t-1}$. Unless $j$ pays the whole wealth $w_{j}^{t\ast\ast}$ and the difference $rw_{i}^{t-1} - w_{j}^{t\ast\ast}$ remains as a ``debt'' that $j$ should repay based on the subsequent production. Since the wealth transfers from $j$ to $i$, the edge weight $p_{ji}$ is added by $\eta$ (\eqref{eq:network_update2}). Finally, the edge weights are normalised so that the sum of edge weights directed from each individual equals to $1$ (\eqref{eq:connectivity_normalisation}).

Then, we arrange the above algorithm for the full model as follows. Each person has the lifetime $l_i$ following the Poisson distribution with the mean $l$. The denominator of productivity would be $l_i$ instead of $100$ in \eqref{eq:production}, to keep the average productivity in a generation fixed.
Then, after the $l_i$ steps pass, each reproduces. Here, the number of children of $i$ follows the Poisson distribution with the mean $1 + \log(1 + w_i)$. Then, each child inherits the parent's wealth and connectivity, i.e., the edge weight directed to the parent, by dividing them equally.

{\small
\section*{Acknowledgements}
The authors thank Heidi Colleran and Tetsuhiro S. Hatakeyama for a stimulating discussion, and Takumi Moriyama, Koji Hukushima, and Takahiro Miyachi for illuminating comments. 
This study was partially supported by a Grant-in-Aid for Scientific Research on Innovative Areas (17H06386) from the Ministry of Education, Culture, Sports, Science, and Technology (MEXT) of Japan. KI is supported by JSPS KAKENHI Grant Number JP21J21565. 
KK is supported by Novo Nordisk Fonden.

\section*{Author contributions statement}
K.I. and K.K. designed the model; K.I. conducted the simulations; K.I. and K.K. analysed the data; and K.I. and K.K. wrote the paper.

\section*{Additional information}
The authors declare no conflict of interest.

Source codes for the model can be found here: \url{https://github.com/KenjiItao/gift.git}
}

\end{document}